\documentclass[letterpaper,journal]{IEEEtran}
\usepackage{amsmath,amsfonts}
\usepackage{array}
\usepackage{textcomp}
\usepackage{stfloats}
\usepackage{url}
\usepackage{hyperref}
\usepackage{verbatim}
\usepackage{graphicx}
\usepackage{balance}
\usepackage{soul,color}
\usepackage{booktabs}
\usepackage{colortbl}
\usepackage[caption=false,font=footnotesize]{subfig}

\newcommand{\gray}[1]{{\textcolor[rgb]{0.5,0.5,0.5}{#1}}}

\begin{document}
\title{When Distributed Consensus Meets Wireless Connected Autonomous Systems: A Review and A DAG-based Approach}
\author{Huanyu Wu, Chentao Yue, ~\IEEEmembership{Member,~IEEE,} Lei Zhang, ~\IEEEmembership{Senior Member,~IEEE,} Yonghui Li, ~\IEEEmembership{Fellow,~IEEE,} and Muhammad Ali Imran,~\IEEEmembership{Fellow,~IEEE}
\thanks{Huanyu Wu, Lei Zhang and Muhammad Ali Imran are with James Watt School of Engineering, University of Glasgow, United Kindom (email: \url{h.wu.3@research.gla.ac.uk}, \url{{Lei.Zhang, Muhammad.Imran}@glasgow.ac.uk)}; Chentao Yue and Yonghui Li are with School of Electrical and Computer Engineering, The University of Sydney, Australia (email: \url{{chentao.yue, yonghui.li}@sydney.edu.au).} }}

\markboth{Journal of \LaTeX\ Class Files,~Vol.~18, No.~9, September~2020}%
{How to Use the IEEEtran \LaTeX \ Templates} 

\maketitle

\begin{abstract}

The connected and autonomous systems (CAS) and auto-driving era is coming into our life. To support CAS applications such as AI-driven decision-making and blockchain-based smart data management platform, data and message exchange/dissemination is a fundamental element. The distributed message broadcast and forward protocols in CAS, such as vehicular ad hoc networks (VANET), can suffer from significant message loss and uncertain transmission delay, and faulty nodes might disseminate fake messages to confuse the network. Therefore, the consensus mechanism is essential in CAS with distributed structure to guaranteed correct nodes agree on the same parameter and reach consistency. 
However, due to the wireless nature of CAS, traditional consensus cannot be directly deployed. This article reviews several existing consensus mechanisms, including average/maximum/minimum estimation consensus mechanisms that apply on quantity, Byzantine fault tolerance consensus for request, state machine replication (SMR) and blockchain, as well as their implementations in CAS. To deploy wireless-adapted consensus, we propose a Directed Acyclic Graph (DAG)-based message structure to build a non-equivocation data dissemination protocol for CAS, which has resilience against message loss and unpredictable forwarding latency. Finally, we enhance this protocol by developing a two-dimension DAG-based strategy to achieve partial order for blockchain and total order for the distributed service model SMR.

\end{abstract}


\vspace{-0.2cm}
\section{Introduction}
\IEEEPARstart{C}{onnected} Autonomous Systems (CAS) is an emerging technology that is expected to enable collaborative autonomous driving and intelligent transportation systems. CAS has been extensively investigated by researchers in recent years, due to the development of protocols such as the emerging vehicular ad hoc network (VANET) as well as the rapid growth of 5G hardware infrastructures including Road Side Unit (RSU) \cite{mollah2020blockchain}. To achieve an intelligent autonomous system, collecting and processing massive data to make \gray{the }optimal and reliable decisions quickly is an inevitable requirement. This requirement can be fulfilled through information sharing and distributed collaborative data processing, supported by distributed techniques including swarm artificial intelligence (AI), distributed data analysis and blockchain~\cite{mollah2020blockchain,xu2022wireless}, etc. 
 
Connectivity among the agents in CAS is the key for collecting and processing massive data. There are two approaches, namely centralised and distributed. 
In centralised CAS approaches, the data from local sensors within a vehicle is either processed locally or sent to the central server, which is responsible for data processing with the result being returned to the vehicles. However, such methods introduce the vulnerability of a single point of failure and potential leaks of sensitive information. Additionally, centralised approaches face substantial overheads and scalability issues as the number of vehicles increases  \cite{feng2023wireless}. 
On the other hand, the central server usually covers fixed geographical areas, relies on vast-amount of distributed real-time data to make critical decisions, and leverages advanced analytics and machine learning algorithms to optimise traffic or predict congestion. Nevertheless, centralised CAS structure is incapable of processing cross-geographical real-time data
where distributed structure could make benefit. As a solution, distributed data process could enable fast response, distribute heavy tasks across multiple nodes and improve security by processing data closer to the source and limiting the dissemination of certain information~\cite{feng2023wireless,arthurs2022taxonomy}.

In distributed CAS approaches, distributed consensus plays an important role either to achieve consistency among different nodes~\cite{cao2022v2v} or to perform as a fundamental protocol to support a trustworthy network~\cite{diallo2021improved}. However, due to the unreliable wireless channels in CAS, the message passing between participants suffers from uncertainty including unpredictable packet loss and transmission delay. The conventional state machine replication (SMR) faulty tolerance techniques, which require the reliable transmission of messages from honest participants, fall short of adapting to such scenarios. For instance, in protocols such as Practical Byzantine Fault-Tolerant (PBFT) consensus \cite{castro1999practical}, the adversary could only delay messages but is unable to intercept them, while unreliable wireless channels can trigger excessive leader elections and lead to system instability. In addition, SMR was designed to process incoming requests in a consistent sequence across all replicas to ensure system reliability and data consistency. This is typically achieved by using round-based consensus mechanisms to reach an agreement between participants. This approach, however, faces challenges in highly dynamic intermittent networks because it prefers honest participants to be always online. To deal with nodes temporarily going offline, SMR must employ an additional synchronisation protocol. In CAS, participants can frequently enter and exit the service range, fail to receive or send some messages. In this case, synchronisation protocols could be frequently triggered and result in high complexity. Apart from SMR, other distributed consensus includes the average/max./min. value estimating consensus built for wireless sensor networks (WSNs)~\cite{chen2014wireless}. Nevertheless, these value estimation consensus protocols are not suitable for agreements on consecutive decisions/operations in CAS, due to their long convergence time and the nature of value-based data.

The promising technology blockchain that supports a trustworthy CAS~\cite{mollah2020blockchain} is also heavily related to consensus. At the same time, the consensus mechanism is not only important in building a secure CAS network, but is also necessary for blockchain layer built atop it. However, current consensus specifically designed for blockchain~\cite{mollah2020blockchain,zhang2023v} in CAS is only suitable for blockchain itself rather than a unified protocol that can be used for general distributed CAS applications and blockchain. Deploying multiple consensus atop CAS network might increase complexity and cause unnecessary overhead.

In this article, we review the existing consensus approaches for CAS, and investigate novel data dissemination and ordering solutions using Directed Acyclic Graph (DAG). First, we provide an overview of consensus protocols, including average/max./min. value estimating, decision-making protocols in SMR and consensus for blockchain. The challenges of implementing these protocols in CAS as well as the accompanying optimisation strategies are accordingly discussed. Then, we propose a DAG-based data dissemination protocol as a promising solution for CAS that adopts the store-carry-forward (SCF) strategy. Finally, we show that the proposed DAG-based data dissemination protocol could be integrated with a DAG-based ordering mechanism to establish an SMR or blockchain.

\vspace{-0.2cm}
\section{Current Consensus in Wireless CAS} \label{Sec::General}

In CAS, the input and output value in consensus could be a quantity (e.g., average speed) or consecutive decisions, (e.g., ``turn left"). The agreement on quantity could be solved by wireless value estimation consensus, while the requirement of reaching agreement on consecutive decisions lead to widely-known SMR. This section briefly reviews the consensus on quantities and decisions, and their wireless extensions and the consensus for blockchain in wireless CAS.

\vspace{-0.2cm}
\subsection{Consensus for Decisions on Quantity}
\label{sect:wirelessquantity}

Distributed average consensus resolves consistency on a quantity, such as vehicle average speed. In wireless networks, it relies on local observations and isotropic information exchange between neighbours. Sensors begin with their initial values, exchanged in ``steps" for comparison. Nodes update estimates using predefined combinations of their values and neighbours' through a predefined linear combination. After a certain number of iterations, all node data converge to a common average value. Similarly, minimum and maximum value consensus can be used, akin to average consensus~\cite{chen2014wireless}.

The wireless average consensus is originally designed for wireless network, and it has also been applied in CAS. For instance, the authors in ~\cite{jala2018coop} considered road conditions and driving environments, treating connected vehicles as Wireless Sensor Networks (WSNs) for information sharing. Although the wireless consensus is originally designed for wireless networks, this approach is only capable of processing quantities rather than consecutive decisions.

\vspace{-.2cm}
\subsection{Traditional Consensus for Decisions on Requests}

The consensus can also decide the sequencing of a series of consecutive requests, e.g., ``turn left$\rightarrow$turn right$\rightarrow$stop''. This leads to the traditional SMR model which execute consecutive request in the same sequence, also known as the Byzantine problem.

\subsubsection{Byzantine Problem}
A general distributed system with $n$ nodes and $f$ faulty nodes faces challenges such as incorrect values, lost or replayed messages, unpredictable latency, and instances of nodes simply stopping working. The consensus dealing with fail/stop failure is known as the \textit{crash fault-tolerance (CFT) consensus}, while those handling Byzantine/arbitrary failure are called \textit{Byzantine fault-tolerance (BFT) consensus}.
The Byzantine problem includes \textit{Byzantine broadcast} and \textit{Byzantine agreement}~\cite{malkhi2019concurrency}. Byzantine broadcast involves a single sender with an initial value, and the goal is for every correct node to either agree on this value if the sender is correct, or arrive at a blank value if the sender is Byzantine. The more generalised form, Byzantine agreement, assumes that each node has an initial value, and all nodes aim to agree on the same valid value. 

\subsubsection{Fault-tolerance SMR}

The most widely-known Byzantine protocols are those designed for SMR. In distributed databases and websites, the core services often operate in a replicated manner for fault tolerance and reliability. SMR is employed to replicate the system state across a set of replicas, ensuring that they maintain a consistent state. Precisely, each request is assigned a sequence number, and every replica executes the requests in the same order in SMR.

CFT consensus for SMR relies on quorum intersection; that is, for each incoming request, a majority (at least $\frac{n}{2}+1$ nodes, i.e., a quorum, of $n$ nodes in total) must receive the request. This mandates that any two sets of requests will overlap at least at one node and at least one node always has access to the most recent system state, resolving potential conflicts between incoming requests. However, CFT only works with fail/stop failure, which means that all behaviours should adhere to the consensus algorithm design except for ceasing to function. One of the widely-used CFTs for SMR is Raft~\cite{ongaro2014search}.

Besides fail/stop failure, nodes might be controlled or attacked by adversaries and behave arbitrarily. This situation necessitates a Byzantine-fault-tolerant consensus. Miguel Castro and Barbara Liskov proposed the well-known PBFT \cite{castro1999practical} in 1999. PBFT also applies the quorum intersection. Nevertheless, different from CFT that only requires at least one correct node in quorum intersection, PBFT needs the number of correct nodes to excel byzantine nodes.  Specifically, to tolerate up to $f$ Byzantine nodes, PBFT requires $2f+1$ correct nodes to advance the consensus. In PBFT, one replica is elected as the leader, who accepts the request from clients, assigns unique sequence numbers to requests, and disseminates them to other replicas during the consensus rounds.

\vspace{-.2cm}
\subsection{SMR Extended to Wireless CAS}
Although SMR protocols were designed for wire systems originally, they have been integrated into wireless systems to solve consistency problems. Yang and Li~\cite{yang2019vehicle} proposed a wireless distributed vehicular network based on P2P network. Considering that the centralised utilisation of on board units (OBUs) and RSUs in certain areas may result in poor scalability and low bandwidth, they proposed a self-organising network with an undirected graph topology.  When a vehicle enters or exits the network, the adjacency matrix representing neighbour information will transition to a new state. The application of PBFT and Raft consensus in wireless autonomous systems was investigated in \cite{xu2022wireless}. It proposed a framework based on PBFT and Raft to test their performance, including reliability, throughput and latency, in wireless CAS. Cao \emph{et al.}~\cite{cao2022v2v} proposed a connected vehicular consensus framework based on Raft, where the consensus is optimised for multi-hop and multi-channel scenarios with dynamic topology and limited communication range. This work can achieve efficient coordination and  negotiation with minimal delay in latency-sensitive driving systems. In \cite{feng2023wireless}, PBFT was utilised to reach an agreement of subsequent actions among vehicles. To adapt to wireless network, \cite{feng2023wireless} applied an improved gossip communication protocol to synchronise failed vehicles. 

These round-based mechanisms are mainly devised for centralised wireless infrastructures; they are not applicable to SCF scheme and decentralised networks with possible message loss. Particularly, packet loss and message delays in wireless communication may cause some nodes to remain stuck in a previous round, while others advance to the next round, ultimately leading to a system-wide halt. Furthermore, when forwarding is needed, vehicles will retain its packet until a suitable forwarder (could be another vehicle) becomes available. However, this strategy could introduce unacceptable communication delay for traditional consensus due to the unpredictable occurrence of forwarders.

\vspace{-.2cm}
\subsection{SMR for Blockchain in Wireless CAS}\label{Sec::Block}

With the nature of decentralisation, immutability and trustworthiness, blockchain is also built atop CAS network to provide trust for distributed control and coordination. 
One of the primary consensus protocols to build blockchain is SMR, which builds consecutive services that each participant executes the same commands (i.e., adding proposed valid blocks one by one) with the same order.

Vehicular network could be regarded as a permissioned network where each participant is registered and authenticated by the authority (e.g., vehicle registration is needed). SMR has been enhanced for vehicular-network permissioned blockchain~\cite{zhang2023v, diallo2021improved}.  Zhang \emph{et al.}~\cite{zhang2023v} designed a blockchain-oriented consensus to address the vehicular BFT problem. Taking into account the intermittent connectivity of vehicles, their approach ensures that consensus can be executed without interruption during membership changes. Diallo \emph{et al.}  ~\cite{diallo2021improved} proposed a PBFT-based consensus for VANET. Blockchain is deployed among RSUs with a dynamic participants selection mechanism, which only selects participants physically close to RSUs to perform consensus. These approaches are optimisations of consensus deployment that could address the geographical or structure requirement in CAS, keeping the core of traditional SMR protocol. The message loss and forward in the underlying communication layer is not considered.

\vspace{-.2cm}
\subsection{Proof-based Consensus for Blockchain in Wireless CAS}
Besides SMR, the other mechanism for building blockchain is to deploy repeated consensus such as Proof-of-X (PoX) to reach an agreement on which block(s) should be added each time. In CAS, messages may be carried and transmitted by third parties. In contrast, blockchain is generally regarded as a peer-to-peer network. To handle this challenge, Ayaz \emph{et al.}~\cite{ayaz2021proof} proposed a new Proof-of-Quality-Factor (PoQF) consensus with data dissemination, which divides the devices in network into RSUs and vehicles. 
PoQF uses signal-to-interference-noise ratio (SINR) and physical distance as quality factors to identify the reliability of the blocks sent by each message originator. Each vehicle will carry-forward the messages to the next hop closest to it for message dissemination. Votes and quality factors will also be disseminated or forwarded to an RSU. After vote dissemination, every vehicle finally reaches the consensus on the true messages with enough votes. However, a primary limitation of PoX is that it was originally designed for cryptocurrency, hindering its application to general CAS use cases. 

\section{A DAG-based Data Dissemination Mechanism for Decentralised CAS with VANET} \label{Sec::DD}

\begin{figure*}[htbp]
	\setlength{\belowcaptionskip}{-0.4cm}   %
	\centerline{\includegraphics[width=0.72\textwidth, trim=670 385 885 950, clip]{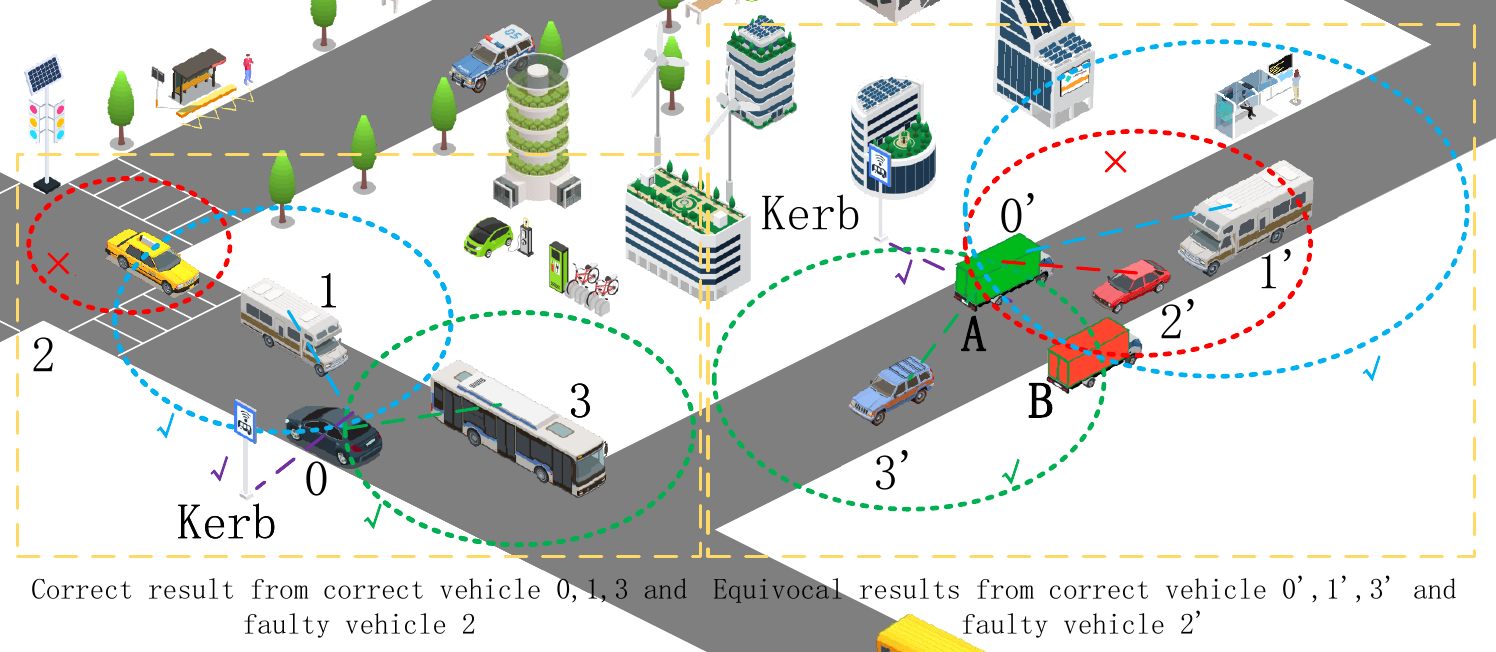}}
	\caption{An example of vehicle location on city roads via CAS.}
	\label{fig:Inconsistency}
\end{figure*}

\subsection{Assumptions for Adapting Consensus for Wireless CAS}
In section \ref{Sec::General} we discussed current widely-used SMR and proof-based consensus in CAS, and identified that SMR is suitable for reaching consensus on a decision of a series of consecutive requests, while proof-based consensus is only suitable for blockchain in CAS. However, the communication in CAS such as VANET suffers from message lose and unpredictable message forward latency, which makes SMR incapable to be directly deployed. In addition, although SMR is the gold standard for implementing distributed functionality, it is in essential a total ordering mechanism, which cannot reach the optimal bound if directly used in CAS scenarios where some decisions need to be ordered, but not all the decisions do. In this section, we assume that correct nodes will broadcast its own correct message or other messages it is responsible for forwarding to other nodes. The messages cannot be forged, yet they might be lost during the broadcasting process. 
A node could crash if it fail/stop working, suffering from sensor data accuracy and distortion problem, but its behaviour is benign.
In addition, Byzantine nodes may appear in the system, which can either fail/stop, broadcast incorrect data or broadcast equivocal data, i.e., sending different data to different nodes.
\subsection{Necessity of Achieving Non-equivocation}
\label{sect:example}

Due to stability and accuracy issues, the sensor could fail/stop and the observed data might be incorrect. These failures could be regarded as crash faults in CAS. Such failure can be found and recovered by cross-validated through information fusion and sharing.
However, a Byzantine node that behaves arbitrarily not only sends incorrect data, but could send equivocal information.
We consider a widely-used triangulation example on city roads to explain crash faults in CAS and equivocation by Byzantine node in Fig. \ref{fig:Inconsistency}. We consider four vehicles $\{0,1,2,3\}$ equipped with sensors $\{a,b,c,d\}$ capable of determining their distance to kerbs or adjacent vehicles.  We assume that vehicle 0 will be positioned. Sensor $b$ measures the distance between vehicles 0 and 1, denoted by $d(0,1)$, while sensors $c$ and $d$ measure the distances $d(0,2)$ and $d(0,3)$, respectively. Sensor $a$ monitors the distance between vehicle 0 and the right kerb $d(0,\mathrm{kerb})$. All data will be shared among the four vehicles for data fusion and process to make a decision. In Fig. \ref{fig:Inconsistency}, the purple, blue and green circles illustrate the correct data obtained by sensors $a$, $b$ and $d$, respectively. The red circle indicates incorrect data from sensor $c$, which is a faulty node that might even be Byzantine.

As illustrated by the left portion of Fig. \ref{fig:Inconsistency}, three distances, e.g., $d(0,1)$, $d(0,3)$, and $d(0,\mathrm{kerb})$, will be sufficient for determining the location of vehicle 0 in triangulation. Vehicles can successfully identify that sensor $c$ provides the faulty data $d(0,2)$ (red circle) through data fusion because the red circle does not intersect with any other pair of circles. On the other hand, the right portion of Fig. \ref{fig:Inconsistency} shows an example of possible equivocation. Vehicle $0'$ will be determined at position A by using the data $d(0',1')$, $d(0',2')$, and $d(0',3')$, while it will be considered at position B with data $d(0',1')$, $d(0',3')$, and $d(0',\mathrm{kerb})$. If $c$ is crash node which captures incorrect data but tells everyone its observation, every participant will consistently decide on gathering further information because they cannot distinguish which position is correct. However, if $c$ is Byzantine which could transmit different data to different participant vehicles (i.e., equivocation), it could result in inconsistent decisions, with some vehicles deciding A and others deciding B. Such inconsistencies could lead to accidents.

Once non-equivocation is achieved, the power of Byzantine node is restrained: it could only broadcast identical incorrect values, otherwise, its equivocal data will be detected and rejected, which results in the identical consequence of fail/stop. In other words, with non-equivocation, the behaviour of Byzantine node is degraded to a crash node. The ordering process is separated from the consensus. Once non-equivocation data dissemination is done, the data which does not need ordering could be directly used, and the data that requires ordering will be further sequenced by the process described in Section\ref{sect:extendtoSMR}.

\vspace{-0.2cm}
\subsection{Overview of the Proposed DAG-based Non-Equivocation Data Dissemination}

BFT can solve the Byzantine problems including equivocation issues mentioned above under the reliable channel assumption. However, in CAS, not only the attackers can spread equivocal information, but the correct data from genuine nodes can be intercepted, jammed and lost due to the unreliable wireless channel. In addition, BFT protocols are monolithic, which achieves non-equivocation based on a critical path of ordering. Nonetheless, as mentioned, not all the data needs to be ordered in CAS.

We proposed a DAG-based data dissemination mechanism to address the issue of packet loss and message delay in CAS, as well as separate non-equivocal data dissemination from the ordering. Precisely, it reduces the number of messages required to be transmitted in each send-receive round, while augmenting the amount of information by including the communication history in each message. The DAG model is employed to construct a shared communication history among participants in order to ensure consistency. %

The proposed protocol is depicted in Fig.~\ref{fig:DAG1}. Specifically, each column of nodes represents a round of data dissemination, during which each vehicle broadcasts its own message and any messages it is supposed to forward. Each row of nodes represents the state progression of a participant. Two nodes are connected if the message is successfully transmitted from one to another. For clarity of Fig.~\ref{fig:DAG1}, the links between nodes are faded if the message is successfully transmitted without loss or delay. 
After the initial round, each participant maintains a local DAG, encompassing all nodes from previous rounds that are connected (either directly or indirectly) to its nodes in the current round. In other words, the DAG captures the communication history related to the node. At each round, participants will update their local DAGs and subsequently broadcast them to other participants.

\subsection{Completeness of Communication History}

The purpose of applying local DAG is to reconstruct a complete communication history for each participant, regardless of message loss and delay. That is, by sharing local DAGs with others, every participant will eventually gain a complete view of the communication history of the first two rounds of all participants, which is referred to as the \textit{history completeness}. Achieving history completeness is sufficient for identifying the equivocation and byzantine nodes.

\begin{figure}[!t]
    \centering
    \subfloat[Round-based dissemination procedure]{\includegraphics[width=0.35\textwidth]{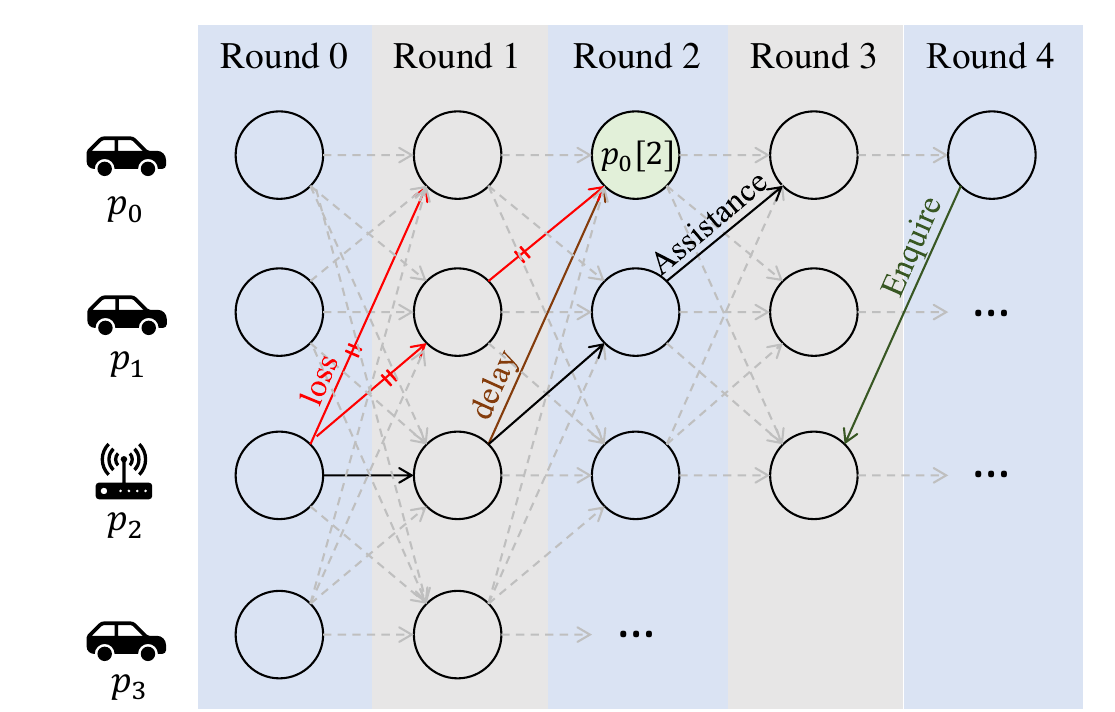}\label{fig:DAG_byz_a}}
    \hfil
    \subfloat[Example of the local DAG stored at {$p_0[2]$}]{\includegraphics[width=0.28\textwidth]{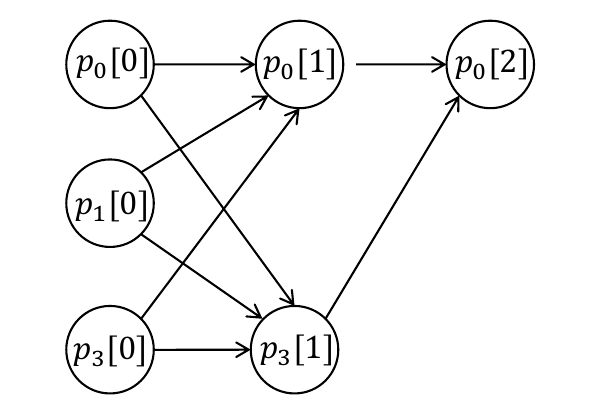}\label{fig:DAG_byz_b}}
    \caption{Illustration of the proposed data dissemination scheme}
    \label{fig:DAG1}
\end{figure}

Denoted by $p_i[r]$ the node of the $i$-th participant at the round $r$, we briefly explain the procedure of achieving history completeness by taking the participant $p_0$ as an example. As shown in Fig.~\ref{fig:DAG1}, node $p_0[2]$ fails to build a path to the node $p_2[0]$ based on its stored local DAG. In this case, $p_0$ will attempt to update its DAG and find a path to $p_2[0]$ in later rounds through methods including retrospection, collaborative assistance, or enquiry.
\begin{itemize}
    \item Retrospection: At round 3, participant $p_0$ may find the delayed message sent from $p_2[1]$ to $p_0[2]$.
    \item Collaborative assistance: Node $p_1[2]$ can share its local DAG to $p_0[3]$ at round 3.
    \item Enquiry: If retrospection and assistance are insufficient to find a path to $p_2[0]$, $p_0$ can make an enquiry to request DAGs from other participants in an additional round.
\end{itemize}
These methods allow $p_0$ to update its local DAG to find a path to $p_2[0]$. With similar approaches, all participants can eventually achieve history completeness, unless the messages from a particular participant are completely lost across all rounds (e.g., leaving the local network/network partition).

\begin{figure}[t]
		\setlength{\abovecaptionskip}{-0.0cm}   %
	\setlength{\belowcaptionskip}{-0.4cm}   %
	\centerline{\includegraphics[width=0.35\textwidth]{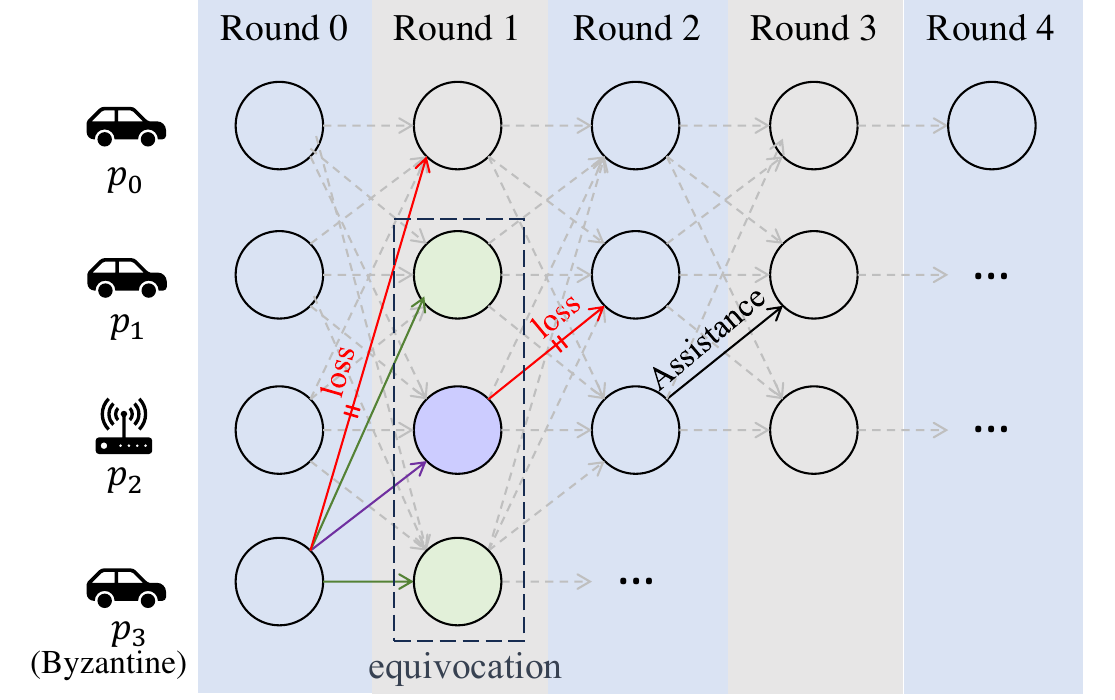}}
	\caption{Inconsistency detection of the proposed mechanism}
	\label{fig:DAG_byz}
\end{figure}

\subsection{Detection of Equivocation}

 As demonstrated in Fig.~\ref{fig:DAG_byz}, we assume that $p_3$ is a Byzantine participant and it broadcasts different values/decisions to $p_1$ and $p_2$, thereby causing equivocation. Participant $p_2$ can easily identify $p_3$ as Byzantine when it advances to round 2, because the message from $p_1[1]$ to $p_2[2]$ will be discovered to be inconsistent with the one it received at $p_2[1]$. However, $p_1$ will fail to detect the equivocation at round 2 due to a packet loss from $p_2[1]$ to $p_1[2]$. In this case, it will try to update its local DAG through retrospection, collaborative assistance, or enquiry. For example, the message from $p_2[2]$ to $p_1[3]$ will assist $p_1$ to update its local DAG and achieve history completeness. Thus, $p_1$ will also be aware of the equivocation caused by $p_3$ at round 3.

Once a participant detects an instance of equivocation, it will invalidate the data from the Byzantine participant, i.e., treating it as NULL, to prevent the propagation of malicious information. Since every correct participant eventually achieves history completeness, the misbehaviour of a Byzantine node will inevitably be detected by all participants. To allow for a maximum of $f$ Byzantine nodes in the proposed protocol, every node can only advance to the next round if it can extract $2f+1$ original messages (i.e., those sent at the first round) from its local DAG.

\subsection{Simulation Result}
We validate the performance of the proposed via simulations\footnote{Code provided on GitHub: https://github.com/BrokenRondo/VANET-DAG-Sim} with different numbers of participants, as recorded in Table.~\ref{tab:1}. We measure the performance by the maximum tolerable loss proportions, defined as the maximum proportion of lost messages to total messages during data dissemination that can be tolerated by the proposed protocol. Note that this concept is different from what is commonly assumed in wireless communication, i.e., each message will follow a certain probability to be lost. However, the high maximum loss proportion still indicates the performance robustness of the protocol in real-world wireless settings. As shown, the tolerable loss proportion increases from 35\% to 80\% when the number of participants rises from 4 to 20, reflecting the performance resilience against packet loss of the proposed protocol. In contrast, the conventional consensus from the literature could halt if more than $\frac{n}{3}$ messages are lost in any round as they tolerate up to 
 $\frac{n}{3}$ faulty nodes~\cite{malkhi2019concurrency}. Table.~\ref{tab:1} also shows the average latency at the maximum tolerable loss proportions, which is measured by the time duration from starting the data dissemination and achieving the history completeness. As seen, the average latency grows with the number of participants.

\begin{table*}[htbp]

  \centering
   \caption{Simulation Result}
  \scalebox{0.80}{
    \begin{tabular}{|>{\columncolor[rgb]{ .949,  .949,  .949}}p{10em} >{\columncolor[rgb]{ .839,  .863,  .894}}rrrr>{\columncolor[rgb]{ .839,  .863,  .894}}rrrr>{\columncolor[rgb]{ .839,  .863,  .894}}rrrr>{\columncolor[rgb]{ .839,  .863,  .894}}r>{\columncolor[rgb]{ .839,  .863,  .894}}r>{\columncolor[rgb]{ .839,  .863,  .894}}r>{\columncolor[rgb]{ .839,  .863,  .894}}r>{\columncolor[rgb]{ .839,  .863,  .894}}r|}
    \hline
    \rowcolor[rgb]{ .949,  .949,  .949}Number of pariticpants & 4     & 5     & 6     & 7     & 8     & 9     & 10    & 11    & 12    & 13    & 14    & 15    & 16    & 20    & 24    & 28    & 32 \\
    Tolerable loss proportion & 0.353  & 0.362  & 0.396  & 0.444  & 0.460  & 0.558  & 0.563  & 0.565  & 0.602   & 0.631  & 0.674  & 0.676  & 0.740  & 0.792 & 0.796 & 0.797 & 0.797 \\
    Average latency (ms) & 105   & 155   & 226   & 328   & 380   & 505   & 739   & 956   & 1358  & 1685  & 1984  & 2645  & 2699  & 3057  & 4380  & 4520  & 5231 \\
   \hline
    \end{tabular}%
    }

  \label{tab:1}%
\end{table*}%

\begin{figure*}[ht]
	\vspace{-0.1cm}  %
		\setlength{\abovecaptionskip}{-0.2em}   %
	\setlength{\belowcaptionskip}{-0.4em}   %
	\centerline{\includegraphics[width=0.75\textwidth]{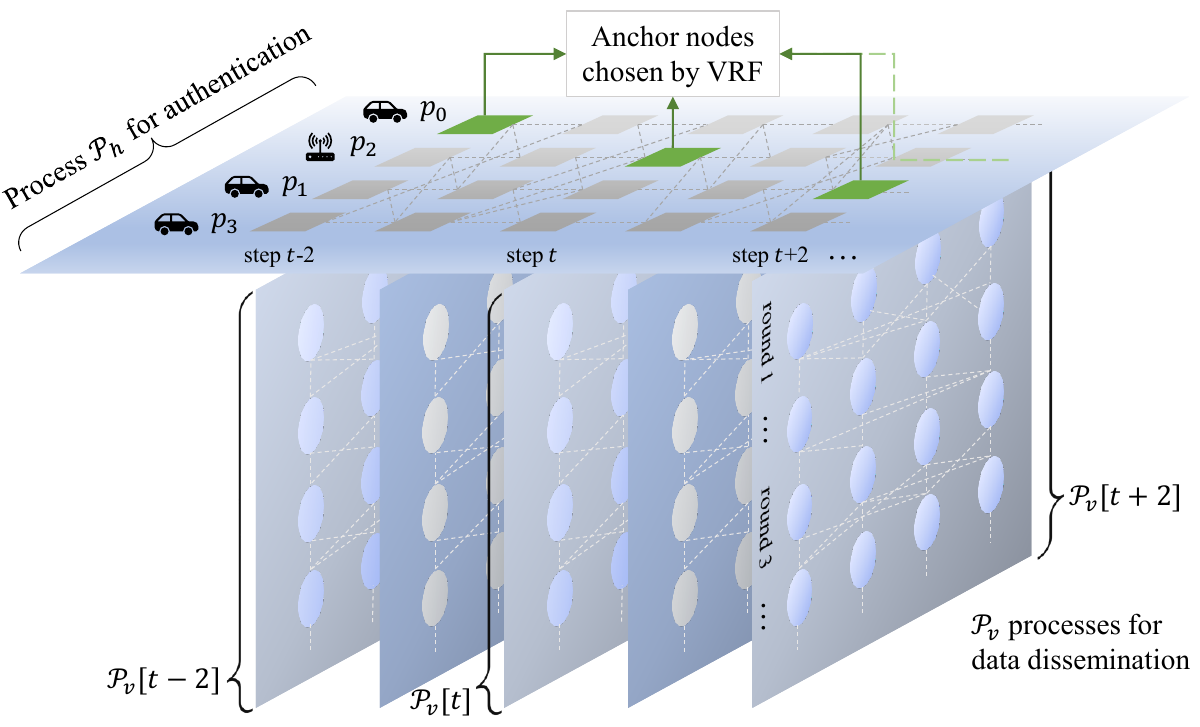}}
	\caption{Two-Dimensional DAG-based Structure}
	\label{fig:DAG_order}
\end{figure*}

\section{Extend to Blockchain/SMR by Ordering}
\label{sect:extendtoSMR}
The autonomous CAS network is expected to be capable of adaptively establishing trust between network entities and protecting the integrity of data exchanged. This can be supported by blockchain and SMR systems with operations/transactions containing application-level information. In both blockchain and SMR, each transaction requires the \textit{ordering}, i.e., the requirement that all operations must be executed in the same order by all network nodes. In CAS, due to the channel noise, possible re-transmission requests, and multiple-access collisions, each transmitted packet will have varying transmission latency, which complicates achieving a consistent and accurate transaction order across all participants. To potentially address this issue, we adopt the availability certificate concept on DAG from Narwhal and Tusk ~\cite{danezis2022narwhal}, and proposed a two-dimensional DAG-based structure for enabling blockchain and SMR. 

\subsection{The Two-Dimensional DAG-based Structure}

As shown in Fig.~\ref{fig:DAG_order}, the proposed scheme is composed of two dimensions of processes: 1)  a horizontal process, denoted by $\mathcal{P}_{h}$ and 2) a series of vertical processes, denoted by $\mathcal{P}_{v}[t]$ for $t = 0,\ldots,t_{\max}$. Specifically, $\mathcal{P}_{h}$ will build DAGs to process the authentication/certificate among nodes, while each $\mathcal{P}_{v}[t]$ runs a DAG-based data dissemination protocol as introduced in Section \ref{Sec::DD}. The process $\mathcal{P}_{h}$ is operated in a step-by-step manner. The step-$t$ nodes of $\mathcal{P}_{h}$ are exactly the round-0 nodes of $\mathcal{P}_{v}[t]$. Processes $\mathcal{P}_{h}$ and $\mathcal{P}_{v}[t]$ are cooperated as follows. 

\begin{itemize}
    \item After the first global stabilisation time (GST), all participants perform the data dissemination protocol, i.e. $\mathcal{P}_{v}[0]$, to broadcast their transactions to others.
    \item If a participant receives at least $n-f$ transactions from others in $\mathcal{P}_{v}[0]$, it performs authentication for these transactions. 
    \item Each authentication forms an edge from a step-$0$ node to a step-$1$ node on $\mathcal{P}_{h}$
\end{itemize}
Recursively, all participants perform the data dissemination protocol $\mathcal{P}_{v}[t]$ to broadcast their transactions, execute authentication for received transactions, and correspondingly update the edges on $\mathcal{P}_{h}$. Then, participants include their messages and the previous authentication results in new transactions and broadcast them in $\mathcal{P}_{v}[t+1]$.

Similar to the protocol introduced in Section \ref{Sec::DD}, each participant will maintain a DAG during $\mathcal{P}_{h}$, which records the history of authentication among participants observed by itself or informed by others. We note that in $\mathcal{P}_{h}$, each participant can only advance to the next step if it receives at least $2f+1$ authentications from the previous step. At early steps, different participants may maintain distinct authentication histories due to packet loss and delay. However, thanks to the dissemination protocol $\mathcal{P}_{v}[t]$ performed at each step, all participants will eventually obtain the same authentication history. In other words, their maintained DAGs in $\mathcal{P}_{v}$ will be updated continuously until eventual consistency.

\subsection{Generation Anchor Vertices for Blockchain}

 A series of anchor nodes need to be selected during the process $\mathcal{P}_{h}$ to construct the blockchain. To ensure that anchor nodes can defend against attacks from malicious participants, an anchor node is chosen for every two steps of $\mathcal{P}_{h}$. Particularly, nodes at step $t+2$ of $\mathcal{P}_{h}$ will determine an anchor node at step $t$. Each anchor node must have at least $f+1$ connections to nodes at the next step to be committed. Since each node maintains a minimum of $2f+1$ edges to the preceding step, the overlap between any $f+1$ and $2f+1$ edges ensures at least one path to the anchor vertex for all correct vehicles. Such anchor nodes can be selected with the approaches of verifiable random function (VRF) and secret sharing~\cite{danezis2022narwhal}. Once an anchor block is committed, the protocol commits all its related authentication history to the blockchain. It has been proved that the worst-case probability to elect a valid anchor block for a certain step is $1/3$~\cite{danezis2022narwhal}. 

\subsection{Partial Ordering and Total Ordering for Blockchain/SMR}

Partial ordering is the ordering of only the anchor nodes in $\mathcal{P}_h$. The partial ordering suffices for the construction of a blockchain. For instance, a ``block'' (i.e., anchor node) can bundle all the nodes between two committed anchor nodes, irrespective of their total order. The partial ordering is achieved by following the links between anchor nodes in $\mathcal{P}_h$, which is readily feasible because links among nodes have been recorded by DAGs. As mentioned, an anchor node requires at least $f+1$ connections to the next step to be committed. However, if an anchor node has less than $f+1$ connections, it will be simply skipped in the ordering. For SMR, the total order is required, which means that all nodes, besides the anchor nodes, need to be ordered according to some metrics. The total order of nodes in $\mathcal{P}_{h}$ can be achieved with a topological sorting algorithm deployed on the subgraph between two anchor nodes. Therefore, through the partial ordering of anchor nodes in $\mathcal{P}_h$, we can efficiently construct blockchains, and with topological sorting, we achieve the necessary total ordering for SMR. This approach forms a versatile consensus mechanism suitable for the unique requirements of both blockchain and SMR systems.

\section{Open Challenges}
The proposed DAG-based scheme can accommodate a forwarding mechanism in CAS and manage message loss. To achieve a fully adaptive consensus protocol for future dynamic CAS, we further identify a few future open challenges.

\subsubsection{Improved low latency}
The next-generation CAS is featured with ultra-reliable low-latency communication (URLLC) \cite{feng2023wireless}. However, when packet loss and message delay are present, our proposed DAG-based solution may require additional rounds (for example, the enquiry action) to achieve non-equivocation, resulting in high processing latency. Thus,
It is essential to investigate novel timeout mechanisms for limiting protocol processing latency in accordance with the application budget, and achieving the best trade-off between latency and system robustness.

\subsubsection{Dynamic configuration}
In CAS, participants could leave the local network without observation and directly cause some vehicles cannot receive enough messages to advance the protocol. Newly joined participants could bring new states required to be updated in DAGs (e.g., adding a new node row in Fig. \ref{fig:DAG_byz_a} during the consensus). Hence, a novel seamless synchronisation and reconfiguration mechanism is required to make the protocol fully adaptive to dynamic networks.

\subsubsection{Garbage Collection}
In the proposed protocol, the DAG is an ever-increasing memory pool, which cannot be stored indefinitely in a vehicle. One can apply the snapshot technique to the protocol, to solidify the common history that has been committed. Then, a garbage collection scheme is expected to release the old history data from memory.

\vspace{-0.2cm}
\section{Conclusion}
In this article, we reviewed the consensus and their implementation in wireless CAS, including average consensus, CFT and BFT as well as blockchain-related consensus. Then, we introduce a DAG-based data dissemination protocol which can guarantee non-equivocation in CAS when packet loss and message delay are present. Furthermore, the proposed DAG-based data dissemination protocol is integrated
with a DAG-based ordering mechanism to enable the application of SMR or blockchain in CAS to ensure trust establishment and data protection.

\ifCLASSOPTIONcaptionsoff
\newpage
\fi
\vspace{-0.3cm}
\bibliographystyle{IEEEtran}
\bibliography{IEEEabrv,ref}

\vskip -2.5\baselineskip plus -1fil

\begin{IEEEbiographynophoto}{Huanyu Wu}
 is pursuing his Ph.D. in Electronics \& Electrical Engineering at James Watt School of Engineering, University of Glasgow (UofG), UK. His research interests include distributed system, blockchain, security and web3.
\end{IEEEbiographynophoto}

\vskip -2.5\baselineskip plus -1fil

\begin{IEEEbiographynophoto}{Chentao Yue}
(Member, IEEE) received B.Sc. in information engineering from Southeast University, China, in 2017, and the Ph.D. degree from the University of Sydney (USYD), Australia, in 2021. He is currently a postdoctoral research associate at the School of Electrical and Computer Engineering (ECE), USYD. 
\end{IEEEbiographynophoto}

\vskip -2.5\baselineskip plus -1fil

\begin{IEEEbiographynophoto}{Lei Zhang}
(Senior muember, IEEE) is a Professor of Trustworthy Systems at UofG. He has academia and industry combined research experience on wireless communications and networks and distributed systems for IoT, and blockchain. 
\end{IEEEbiographynophoto}

\vskip -2.5\baselineskip plus -1fil

\begin{IEEEbiographynophoto}{Yonghui Li} (Fellow, IEEE) is a Professor and Director of Wireless Engineering Laboratory at the ECE, USYD. He is the recipient of the Australian Queen Elizabeth II Fellowship in 2008 and the Australian Future Fellowship in 2012.

\end{IEEEbiographynophoto}

\vskip -2.5\baselineskip plus -1fil
\begin{IEEEbiographynophoto}{Muhammad Ali Imran} (Fellow, IEEE) is a Professor of communication systems with the UofG, UK, and a Dean with Glasgow College UESTC. He is also an Affiliate Professor with the University of Oklahoma, USA, and a Visiting Professor at University of Surrey, UK. 
    
\end{IEEEbiographynophoto}

\end{document}